\documentclass[preprint,12pt]{elsarticle}

\usepackage{color}
\usepackage{amsthm}
\usepackage{amssymb}

\newcommand{\re}{\mbox{$\rm e$}}
\newcommand{\ri}{\mbox{$\rm i$}}

\newcommand{\tr}{\mbox{$\rm tr$}}
\newtheorem{theorem}{Theorem}[]

\journal{Journal of Geometry and Physics}

\begin{document}

\begin{frontmatter}

\title{A Riemannian approach to Randers geodesics}

\author[brunel,pet] {Dorje C. Brody}
\author[cambridge]{Gary W. Gibbons}
\author[brunel]{David M. Meier}

\address[brunel]{Department of Mathematics, Brunel University London, Uxbridge UB8 3PH, UK}
\address[pet]{Department of Optical Physics and Modern Natural Science, 
St Petersburg National Research University of Information Technologies, Mechanics and Optics, 
49 Kronverksky Avenue, St Petersburg 197101, Russia}
\address[cambridge]{Department of Applied Mathematics and Theoretical Physics, Centre for Mathematical 
Sciences, Wilberforce Road, Cambridge CB3 0WA, UK}

\begin{abstract}
In certain circumstances tools of Riemannian geometry are sufficient to address questions arising in the more general Finslerian context. We show that one such instance presents itself in the characterisation of geodesics in Randers spaces of constant flag curvature. To achieve a simple, Riemannian derivation of this special family of curves, we exploit the connection between Randers spaces and the Zermelo problem of time-optimal navigation in the presence of background fields. The characterisation of geodesics is then proven by generalising an intuitive argument developed recently for the solution of the quantum Zermelo problem. 
\end{abstract}


\end{frontmatter}



Investigations of Finsler manifolds usually require tools more involved than those of Riemannian geometry \cite{rund}. 
For instance, whereas the Levi--Civita connection of Riemannian geometry is a linear connection on the tangent bundle 
of the underlying manifold, one of its generalisations in the Finslerian context, the so-called  Chern connection, is a linear connection on a distinguished vector bundle over the projective sphere bundle \cite{Shen_book}. Nevertheless, in certain situations Riemannian methods are sufficient to deal with aspects of Finsler geometry and the 
resulting simplifications, such as the ones reported below, can be substantial. Specifically, what we show is that the main 
result of \cite{robles}, namely, the characterisation of the geodesics
of a special class of Finsler spaces, can be proven using tools from Riemannian geometry only. 

To begin, let us recall that a Finsler manifold $({\mathcal M}, F)$ is a $C^\infty$ manifold ${\mathcal M}$  together with a 
positive function $F(x,y)$ on the tangent bundle, called the Finsler function, which is required to be $C^\infty$ and 
homogeneous of first degree, that is, $F(x, \lambda y) = \lambda F(x, y)$ for any $\lambda > 0$. Moreover, the 
Hessian of $F^2$ with respect to $y$\,:
\begin{eqnarray}
  g_{ij}(x, y) = \frac{1}{2} \frac{\partial^2}{\partial y^i y^j}\, F^2(x, y) 
\nonumber 
\end{eqnarray}
is assumed to be positive-definite outside the zero-section of $T{\mathcal M}$. It can be shown that 
$F(x, y) = \sqrt{g_{ij}(x, y) y^i y^j}$. 

If $F$ can be expressed in the form  
\begin{eqnarray} 
F(x, y) = \sqrt{\alpha_{ij} y^i y^j} + \beta_i y^i, 
\nonumber 
\end{eqnarray} 
where $\alpha$ is a Riemannian metric and $\beta$ a one-form, then ${\mathcal M}$ is called a Randers space. The Finslerian metric on ${\mathcal M}$ of Randers type thus takes the form 
\begin{eqnarray} 
g_{ij}(x,y) = \alpha_{ij} + \beta_i \beta_j + \frac{(\alpha_{ij}\beta_k+\alpha_{jk}\beta_i 
+\alpha_{ki}\beta_j){y}^k}
{(\alpha_{kl}{y}^k{y}^l)^{1/2}} 
+ \frac{(\beta_k y^k) \alpha_{ik}\alpha_{jl} y^k y^l}
{(\alpha_{kl}y^ky^l)^{3/2}} . 
\nonumber 
\end{eqnarray} 
 Randers 
spaces were first introduced in \cite{randers} in the context of a unified theory of gravitation and electromagnetism and 
arise in a wide range of physical applications such as the electron microscope \cite{ingarden}, the propagation of sound 
and light rays in moving media \cite{gibbons2, luneburg}, and the time-optimal control in the presence of background 
fields \cite{shen}---the last point being of particular relevance for the present discussion. 

To explain the connection 
between Randers spaces and time-optimal control, we start from a Riemannian manifold ${\mathcal M}$ with metric $h$, 
together with a vector field $W$ that satisfies $|W| < 1$ and plays the role of background field, or `wind'. The goal is to 
solve the Zermelo problem, that is, to navigate from one point on ${\mathcal M}$ to another along the path $q(s)$ in the 
shortest possible time under the influence of $W$, assuming a maximum attainable speed of $|\dot{q}| = 1$ if wind were 
absent. A problem of this kind was first posed and solved by Zermelo for the navigation of ships at sea (modelled as the 
Euclidean plane) for a general spacetime-dependent field $W$ \cite{zermelo} (see also \cite{caratheodory}). The general 
formulation on Riemannian manifolds under time-independent fields, and the connection to Randers spaces, was identified 
more recently by Shen \cite{shen}. The idea can be illustrated as follows. Supposing for a moment that one were able to 
travel for finite time in a tangent space $T_{p}{\mathcal M}$ for a fixed $p$, it is clear that the set of destinations reachable 
in one unit of time coincides with the unit circle, shifted by $W(p)$. Correspondingly, the minimum time $F(p, v)$ it takes 
to reach the tip of a given vector  $v$  in $T_{p}{\mathcal M}$ is given by the ratio $|v|/ |\rho_v|$ of  Euclidean norms, 
where $\rho_v$ is the unique vector collinear with $v$ that lies on the shifted unit circle. To put it differently, the vector 
$v/F(p, v) - W(p)$ has unit length. It follows that
\begin{eqnarray}
  F(p, v) = \frac{- h(v, W(p)) + \sqrt{h(v, W(p))^2 + |v|^2 (1 - |W(p)|^2)} }{1 - |W(p)|^2}.
\nonumber
\end{eqnarray}
The function $F$ defined in this manner is a Finsler function of Randers type. Specifically, \begin{eqnarray}
  \alpha_{ij} = \frac{h_{ij}}{1 -  |W|^2} + \frac{W_i W_j}{(1 - |W|^2)^2}, \qquad \beta_i = - \frac{W_i}{1 - |W|^2},
\nonumber 
\end{eqnarray}
where $W_i = h_{ij}W^j$. Conversely, it can be shown that for each Randers space there is a corresponding Zermelo 
problem \cite{robles0}. 
We remark in passing that there is yet another equivalent perspective, whereby with each 
Randers space is associated a conformally stationary spacetime \cite{gibbons1}. 

The preceding discussion implies that if a curve $q: [a, b] \to {\mathcal M}$ is traversed at maximum speed, then the 
time it takes to 
complete the journey is given by the Randers length
\begin{eqnarray}
T = \int_a^b F(q(s), {\dot q}(s)) \, {\rm d}s, 
\nonumber 
\end{eqnarray}
where we wrote ${\dot q}(s)$ for the derivative with respect to the curve parameter $s$. If the curve $q(s)$ has the 
physical parameterisation, that is, $q(s)$ corresponds to the location reached by the maximum speed trajectory at 
time $s- a$ after setting off from $q(a)$, then $F(q(s), {\dot q}(s)) = 1$ and $T = b - a$. In other words, curves in the 
physical parameterisation have unit Randers speed and the passage of time is measured by Randers length. As 
a consequence, Randers geodesics in the physical parameterisation correspond to solutions of the Zermelo 
problem. To make this statement more precise, recall that Randers geodesics are curves that locally minimise 
Randers length. That is, $q: [a, b] \to {\mathcal M}$ is a Randers geodesic if and only if for any $c \in [a, b]$ there 
exists an interval $I = [c- \varepsilon, c + \varepsilon]$ such that $q|_I$ minimises Randers length among all 
curves defined on $I$ with the same endpoints. Hence, if endowed with the physical parameterisation, Randers 
geodesics are the same as curves that locally minimise travel time. Using this equivalence, we can reformulate 
Theorem 2 of \cite{robles} in the following equivalent manner. We write $\mathcal{L}$ for Lie derivative. 

\begin{theorem}
Assume that the wind vector field $W$ in the Zermelo problem above is an 
infinitesimal homothety, that is, $\mathcal{L}_W h = \sigma h$ for a constant $\sigma$. Then, if $q : (-\varepsilon, 
\varepsilon) \to {\mathcal M}$ is a locally time-minimising curve, $p(t) = \varphi_t(t, q(t))$ is a Riemannian 
geodesic of $({\mathcal M}, h)$, where $\varphi$ is the flow of $-W$.  Conversely, if $p: (-\varepsilon, \varepsilon) 
\to {\mathcal M}$ is a Riemannian geodesic,  $q(t) = \varphi^{-1}(t, p(t))$ is a locally time-minimising curve, where 
$\varphi^{-1}$ is the flow of $W$. 
\label{thm:1} 
\end{theorem} 

Notice that the existence of the flow maps on neighbourhoods containing $q(t)$ and $p(t)$, respectively, can be 
ensured by scaling $\varepsilon$ if necessary. 

The proof of Theorem~2 of \cite{robles} (reformulated here as Theorem~\ref{thm:1} above) relies on the geodesic 
equation for Randers spaces, derived for instance in \cite[Chapter 11]{Shen_book}, which is then verified by explicit 
calculation---but as we saw above, the Randers geodesics on $({\mathcal M}, F)$ correspond precisely to the locally 
time-minimising curves of the Zermelo problem. To exploit this fact, in the above formulation of the theorem 
we have made direct reference to the solution curves of the Zermelo problem, which suggests that a Riemannian  
proof, without the derivation of the equation characterising Randers geodesics as a prerequisite, should be possible. 
Before we proceed with this, let us remark first that the theorem applies in particular to Randers spaces of constant flag curvature, 
since their wind vector fields are homotheties \cite{robles0}. 

To gain an intuition for our Riemannian 
derivation, it will be instructive to examine a concrete example. For this purpose let us consider a particular problem of 
time-optimal quantum control. In the quantum Zermelo navigation problem, introduced by Russell \& Stepney in 
\cite{stepney2}, one considers a quantum system under the influence of an ambient field characterised by a 
Hamiltonian operator $\hat{H}_0 \in \mathfrak{su}(N)$, whose Hilbert--Schmidt norm $[\tr(H_0^2)]^{1/2}$ is less 
than unity (in a suitable unit of energy). The goal is to find the time-dependent control Hamiltonian ${\hat H}_1(t)$ 
that satisfies the bound $\tr({\hat H}_1(t)^2) \leq 1$ and achieves, in shortest possible time, the transformation 
$\hat{U}_I \to \hat{U}_F$ between specified initial and final unitary operators (quantum gates) in SU($N$). It 
was shown in \cite{stepney3, brody3} that the optimal control Hamiltonian takes the simple form
\begin{eqnarray}
  \hat{H}_1(t) = {\re}^{-{\rm i} \hat{H}_0 t} \hat{H}_1(0)\, {\re}^{ {\rm i}\hat{H}_0 t}.
\nonumber 
\end{eqnarray}
Moreover, the solution $\hat{U}(t)$ of the Schr\"{o}dinger equation 
\begin{eqnarray}
\frac{\rm d}{ {\rm d}t} \hat{U}(t) = - {\rm i} (\hat{H}_0 + \hat{H}_1(t))\, \hat{U}(t) 
\nonumber 
\end{eqnarray}
emanating from $\hat{U}_I$ is given by
\begin{eqnarray}
  \hat{U}(t) = {\re}^{-{\rm i} \hat{H}_0 t}\, {\re}^{-{\rm i} \hat{H}_1(0) t}\, \hat{U}_I  
\nonumber 
\end{eqnarray} 
on account of the special form of ${\hat H}_1(t)$ above, together with standard results in the 
interaction-picture analysis of quantum mechanics \cite{sunakawa}. Alternatively, one can verify by 
differentiation of $\hat{U}(t)$ that the relevant Schr\"{o}dinger equation is indeed satisfied. We now 
develop an intuitive characterisation of the time-optimal solution $\hat{U}(t)$, which will be useful for later analysis. 
For this purpose, we first recast the solution in the form
\begin{eqnarray}
{\re}^{{\rm i} \hat{H}_0 t} \, \hat{U}(t) =  {\re}^{-{\rm i} \hat{H}_1(0) t}\, \hat{U}_I. %
\nonumber 
\end{eqnarray}
 Let us abbreviate the expression on the left hand side by $\hat{Z}(t)$. That is, $\hat{Z}(t)$  represents the curve 
 $\hat{U}(t)$ in a frame that is dragged along by the right-invariant vector field on SU($N$) given by 
 $\hat{W}(\hat{U}) = -\ri \hat{H}_0 \hat{U}$. Clearly, $\hat{Z}(t)$ starts at $\hat{U}_I$ and hits the `moving target' 
 $ {\re}^{{\rm i} \hat{H}_0 t} \hat{U}_F$ at some optimal time $T$.  Moreover, one can check by differentiation that 
\begin{eqnarray} 
  \frac{\rm d}{ {\rm d}t} \hat{Z}(t) = -\ri \, {\re}^{{\rm i} \hat{H}_0 t}\, \hat{H}_1(t)\,  {\re}^{-{\rm i} \hat{H}_0 t} \, \hat{Z}(t). 
\nonumber 
\end{eqnarray} 
This implies that the right-invariant velocity $(\partial_t \hat{Z}(t)) \hat{Z}^{-1}(t)$ has  unit length in the Hilbert--Schmidt 
norm no matter how the control ${\hat H}_1(t)$ is chosen, so long as it satisfies the full-throttle condition $\tr({\hat H}_1(t)^2)=1$. 

To put the matter differently, let us introduce the bi-invariant Riemannian metric $\gamma$ on SU($N$) whose 
restriction to $\mathfrak{su}(N)$ coincides with the Hilbert--Schmidt inner product $\gamma(A, B) = \tr(A, B)$. Then 
the preceding velocity constraint can be expressed in the form $\gamma(\partial_t \hat{Z}(t), \partial_t \hat{Z}(t))=1$, or, 
more succinctly, $|\partial_t \hat{Z}(t)|=1$. Clearly, any optimal control must meet the full-throttle condition at all times 
and will thus indeed satisfy $|\partial_t \hat{Z}(t)|=1$. Hence, since the speed of $\hat{Z}(t)$ is fixed, the only strategy 
for shortening the time until the moving target is intercepted, is to shorten the path that is traversed in the interim. 
Therefore, $\hat{Z}(t)$ should be a Riemannian geodesic on SU($N$). 
Alternatively stated, Randers geodesics in the Schr\"odinger picture should correspond to Riemannian 
geodesics in the interaction picture. This conclusion is indeed borne out by the 
right hand side of the relation ${\re}^{{\rm i} \hat{H}_0 t} \, \hat{U}(t) =  {\re}^{-{\rm i} \hat{H}_1(0) t}\, \hat{U}_I$, and is 
in agreement with Theorem~\ref{thm:1}, upon noting that $\hat{W}$ induces an isometric flow and the parameter 
$\sigma$ appearing in Theorem~\ref{thm:1} thus vanishes. A similar intuitive argument can be developed for the 
time-optimal control of quantum states (rather than gates) in the presence of background fields \cite{BrGiMe2015}. 

A key ingredient in the foregoing example is the method, familiar from mechanics and optimal control theory 
\cite{Jur1, Jur2,  Agrachev}, of switching to a \textit{moving frame}. 
This strategy can be generalised  to the 
generic case considered in Theorem~\ref{thm:1}. To see this, let $C(t)$ be a time-varying control, assumed to satisfy 
$h(C(t), C(t)) = |C(t)|^2 = 1$ at all times. By definition, any trajectory $q(t)$ produced by the control satisfies 
$\partial_t q(t) = W(q(t)) + C(t)$. Writing $\varphi(t, \cdot) = \varphi_t(\cdot)$ for the flow of $-W$ at time $t$, we 
introduce the curve $p(t) = \varphi_t(q(t))$.
Intuitively speaking, $p(t)$ represents $q(t)$ in a coordinate frame that is pulled along by the wind. If $q(t)$ is 
time-optimal between $q(0) = q_I$ and $q(T) = q_F$, then $p(T) = \varphi_T(q_F)$. That is, $p(t)$ intercepts 
the moving target $\varphi_t(q_F)$ in the shortest possible time. Writing $D\varphi_t$ for the differential of 
$\varphi_t$, we find
\begin{eqnarray}
\frac{{\rm d}}{{\rm d}t}\, \varphi_t(q(t)) &=& -W(p(t)) + D\varphi_t(\dot{q}(t)) \nonumber \\ 
&=&  -W(p(t)) + D\varphi_t( W(\varphi_t^{-1}(p(t))) + C(t)) =  D\varphi_t( C(t)),
\nonumber 
\end{eqnarray}
where in the last step we used the fact that (cf. \cite[Proposition 9.41]{lee})
\begin{eqnarray} 
\frac{{\rm d}}{{\rm d}t}\, D\varphi_t(W\circ \varphi_t^{-1}) = D\varphi_t([W, W] \circ \varphi_t^{-1}) = 0
\nonumber 
\end{eqnarray}
and therefore that $D\varphi_t(W \circ \varphi_t^{-1}) = W$ for all $t$. Since by assumption $W$ is an infinitesimal 
homothety, we have for any vector $v$ that
\begin{eqnarray}
\frac{{\rm d}}{{\rm d}t} \, h(D \varphi_t(v), D\varphi_t(v)) = -\sigma \, h(D \varphi_t(v), D\varphi_t(v)) 
\nonumber 
\end{eqnarray}
and hence that $ h(D \varphi_t(v), D\varphi_t(v)) = {\rm e}^{-\sigma t} h(v, v)$. As a consequence we deduce 
that 
\begin{eqnarray}
h(D \varphi_t(C(t)), D\varphi_t(C(t))) =  {\rm e}^{-\sigma t}. 
\nonumber 
\end{eqnarray}
This shows, in particular, that the speed 
$|\partial_t p(t)| = {\rm e}^{-\sigma t/2}$ is independent of the chosen control. It follows that the time until the 
moving target is intercepted is a strictly increasing function of the distance traveled, and intuition thus dictates 
that if $q(t)$ is a time-optimal curve between $q_I$ and $q_F$, then $p(t)$ should be a geodesic. 
To prove Theorem~\ref{thm:1} we only need to make this intuition rigorous. 

\begin{proof} 
Let $q: [-\varepsilon, \varepsilon] 
\to {\mathcal M}$ be locally time-minimising. Then there exists for each $s \in (-\varepsilon, \varepsilon)$ a $\delta >0$ 
such that $q_s(t) = q(s + t)$ solves the Zermelo problem between $q_I = q(s)$ and $q_F = q(s+ \delta)$. Hence, the 
curve $p_s(t) = \varphi_t(q_s(t))$ defined on $[0,\delta]$ (existence of the flow map can be ensured by scaling $\delta$, 
if necessary) has fixed speed, as in the discussion above, and intercepts the moving target $\varphi_t(q_F)$ at the 
shortest possible time $\delta$. It must therefore be a geodesic on $[0, \delta]$, since otherwise one could reach the 
point $\varphi_\delta(q_F)$ before $\delta$ and then navigate towards the target to obtain an earlier rendezvous. But 
clearly, if $p_s(t)$ is a geodesic on $[0, \delta]$, then $p(t)= \varphi_t(q(t)) = \varphi_s(p_s(t-s))$ is a geodesic on 
$[s, s+\delta]$, bearing in mind that $\varphi_s$ scales the metric by a constant. Since $s$ was arbitrary, we conclude 
that $p(t)$ is a geodesic on $(-\varepsilon, \varepsilon)$, as required. 

Conversely, suppose that $p(t) = \varphi_t(q(t))$ is a geodesic on $(-\varepsilon, \varepsilon)$. Just as before, one 
can take any $s \in (-\varepsilon, \varepsilon)$ and find a $\delta>0$ such that $p_s(t) = p(s + t)$ is length-minimising 
between $p(s)$ and $p(s+ \delta)$. The claim is that $q_s(t) = q(s + t)$ is a time-minimising trajectory between 
$q_s(0) = q(s)$ and $q_s( \delta) = q(s + \delta)$. Suppose this were not the case. That is, there exists a curve $Q(t)$ 
commencing at $q(s)$ and arriving at $q(s + \delta)$ at time $\delta' < \delta$. If we set $P(t) = \varphi_t(Q(t))$ for 
$t \leq \delta'$ then our earlier calculations show that this curve commences at $q(s)$, has speed ${\re}^{-\sigma t}$ 
and intercepts the moving target $\varphi_t(q(s+\delta))$ at time $\delta' < \delta$. For $t > \delta'$, define $P(t)$ to move 
from the interception point along the flow line of $-W$, its speed still satisfying the same constraint. By choosing 
$\delta$ small enough, one can ensure that $P(t)$ moves at above wind speed and thus arrives, still before time 
$\delta$, at $\varphi_\delta(q(s + \delta))$. Observing that $P(t)$ and $\varphi_s^{-1}p_s(t) = \varphi_t(q(s+t))$ have 
the same speed and that the latter curve takes a longer time, $\delta$, to arrive at  $\varphi_\delta(q(s + \delta))$, 
one concludes that the distance travelled by $P(t)$ must be shorter. Upon applying $\varphi_s$ to both paths, this 
implies that $\varphi_s(P(t))$, connecting $p_s(0) = p(s)$ and $\varphi_{s + \delta} (q(s + \delta)) = p_s(\delta)$, is 
shorter than $p_s(t)$, in contradiction to the length-minimising property of the latter curve. This completes the proof 
of Theorem~\ref{thm:1}.
\end{proof} 
\vspace{1mm}
To summarise, we have presented a novel treatment of geodesics in a certain class of Randers spaces that 
includes those of constant flag curvatures. Taking 
inspiration from a recent analysis of the quantum Zermelo problem in the interaction-picture, 
we achieved a formulation that relied exclusively on standard Riemannian methods, thus bypassing a more involved 
Finslerian analysis of Randers spaces. Simplifications of this nature serve an important purpose in making 
results that have been formulated in a specialised mathematical language accessible to a wider audience. At the 
same time our analysis offers an instance whereby the strength of physical intuition contributes to transparency in 
abstract mathematical reasoning. As a final remark we mention that Randers geodesics in spaces of constant flag 
curvature correspond to null geodesics of conformally flat spacetimes \cite{gibbons1}. It would be interesting to 
analyse Theorem~\ref{thm:1} in this spacetime context.






\end{document}